\documentclass[acmsmall]{acmart}

\usepackage{soul}
\usepackage{multirow}
\usepackage{dcolumn}
\usepackage{enumitem}
\usepackage{balance}
\usepackage{booktabs}
\usepackage{graphicx}
\usepackage{comment}
\usepackage{adjustbox}

\AtBeginDocument{%
  \providecommand\BibTeX{{%
    \normalfont B\kern-0.5em{\scshape i\kern-0.25em b}\kern-0.8em\TeX}}}

\usepackage{tabularx}
\usepackage{colortbl}
\usepackage{ifthen}
\usepackage{bm}
\usepackage{lscape}
\usepackage{afterpage}

\usepackage{dcolumn}
\newcolumntype{d}[1]{D{.}{.}{#1}}

\definecolor{cExcellent}{HTML}{31A354}
\definecolor{cGood}{HTML}{6DBF6F}
\definecolor{cModerate}{HTML}{C1EBBC}
\definecolor{cPoor}{HTML}{FFED6F}
\definecolor{cVpoor}{HTML}{FFC300} 

\newcommand{\icc}[1]{%
    \cellcolor{cExcellent}%
    \ifthenelse{\lengthtest{#1 pt < 0.90pt}}{\cellcolor{cGood}}{}%
    \ifthenelse{\lengthtest{#1 pt < 0.75pt}}{\cellcolor{cModerate}}{}%
    \ifthenelse{\lengthtest{#1 pt < 0.5pt}}{\cellcolor{cPoor}}{}%
    \ifthenelse{\lengthtest{#1 pt < 0.25pt}}{\cellcolor{cVpoor}}{}%
    #1%
}

\setcopyright{acmlicensed}
\acmJournal{PACMHCI}
\acmYear{2024} \acmVolume{8} \acmNumber{CHI PLAY} \acmArticle{293} \acmMonth{10}\acmDOI{10.1145/3677058}

\begin{document}

\title{An Investigation of the Test-Retest Reliability of the miniPXI}

\author{Aqeel Haider}
\email{aqeel.haider@kuleuven.be}
\affiliation{%
  \institution{KU Leuven}
  \country{Belgium}}
  
\author{Günter Wallner}
\email{guenter.wallner@jku.at}
\affiliation{%
  \institution{Johannes Kepler University Linz}
  \country{Austria}}

\author{Kathrin Gerling}
\affiliation{%
  \institution{Karlsruhe Institute of Technology}
  \country{Germany}}
\email{kathrin.gerling@kit.edu}

\author{Vero Vanden Abeele}
\affiliation{%
  \institution{KU Leuven}
  \country{Belgium}}
\email{vero.vandenabeele@kuleuven.be}

\renewcommand{\shortauthors}{Haider et al.}

\begin{abstract}
Repeated measurements of player experience are crucial in games user research, assessing how different designs evolve over time. However, this necessitates lightweight measurement instruments that are fit for the purpose. In this study, we conduct an examination of the test-retest reliability of the \emph{miniPXI}---a short variant of the \emph{Player Experience Inventory} (\emph{PXI}), an established measure for measuring player experience. We analyzed test-retest reliability by leveraging four games involving 100 participants, comparing it with four established multi-item measures and single-item indicators such as the Net Promoter Score (\emph{NPS}) and overall enjoyment. The findings show mixed outcomes; the \emph{miniPXI} demonstrated varying levels of test-retest reliability. Some constructs showed good to moderate reliability, while others were less consistent. On the other hand, multi-item measures exhibited moderate to good test-retest reliability, demonstrating their effectiveness in measuring player experiences over time. Additionally, the employed single-item indicators (\emph{NPS} and overall enjoyment) demonstrated good reliability. The results of our study highlight the complexity of player experience evaluations over time, utilizing single and multiple items per construct measures. We conclude that single-item measures may not be appropriate for long-term investigations of more complex PX dimensions and provide practical considerations for the applicability of such measures in repeated measurements.
\end{abstract}



\maketitle

\section{Introduction}
    \label{sec:introduction}

Games user research (GUR) plays a key role in evaluating player experience (PX). Through collecting empirical data via user evaluations, GUR offers valuable insights for game development, aspiring to measure and predict success indicators such as player engagement or perceived enjoyment~\cite{bernhaupt_game_2015}. In the past decade, the methodologies employed in GUR have undergone rapid advancements, from playability heuristics~\cite{Desurvire_heuristics_2009} to rapid iterative evaluations~\cite{pagulayan_user-centered_2003}, from biometric data collection~\cite{pejman_biometrics_2012} to advanced machine learning and adaptive AI~\cite{hsieh2021adaptive}. However, despite these innovations in GUR through objective data and analytics, gathering subjective evaluations of PX remains a cornerstone of GUR. In consequence, questionnaires continue to be a crucial element of GUR, serving as a means of capturing self-reported PX, including aspects such as overall enjoyment, mastery, or immersion~\cite{josefPX2016}. 

From a methodological point of view, preferred measures for evaluating PX include multiple dimensions and multiple items per dimension, thus providing higher reliability, which is crucial for the validity of a psychometric instrument~\cite{nunnally_1978}. To this end, a number of measures containing multiple dimensions and items to evaluate PX have been developed and validated~\cite{BROCKMYER_GEQ_2009,cheng_giq_2015, Jennet_IEQ_2008,Ryan_PENS_2006,azadvar_upeq_2018,Abeele2020}. However, participants often perceive multidimensional measures as lengthy, resulting in fatigue and loss of attention, which, in turn, hinders their widespread adoption~\cite{haider_minipxi_2022}. The aforementioned drawbacks are particularly prominent in the context of rapid iteration cycles and tight timelines common in game development and in settings that require repeated measurement of PX in combination with other measures.

To address participant fatigue and shorten the time needed for GUR evaluations, shorter measures have recently emerged, such as the \emph{HEXAD-12} ---a short version of the \emph{Gamification User Types Hexad Scale}~\cite{krath_hexad12_2023} and the \emph{GUESS-18} ---a shorter version of the \emph{Game User Experience Satisfaction Scale} (\emph{GUESS})~\cite{keebler_GUESS_2020}, both using fewer items per PX construct. This has practical advantages such as quicker response times~\cite{nagy_time_2002}, lower cost~\cite{Wanous1997}, higher face validity~\cite{Dolbier_2004}, lower respondent dissatisfaction~\cite{stanton_development_2002}, and fewer data omissions~\cite{Cheon2015}. However, while these metrics offer advantages in terms of efficiency and convenience, their reliability and validity when capturing the multifaceted nature of PX requires additional investigation~\cite{haider_minipxi_2022}. In particular, further research is needed to verify the extent to which single-item per PX construct or shorter variants using fewer items per PX construct can effectively capture the diverse and complex dimensions of PX and the extent to which they are stable over a longer period of time. 

We address this issue leveraging the \emph{miniPXI}~\cite{haider_minipxi_2022}, a recently developed short version of the \emph{PXI}~\cite{Abeele2020}. Prior validation of the \emph{miniPXI} demonstrated that it can be a valuable tool for PX evaluations where longer measurements are not feasible. The development and validation of the \emph{miniPXI} was either through delayed recall of game experiences or via the sensitivity analysis of fully developed games with a limited sample~\cite{haider_minipxi_2022}. 
 
Additionally, a preliminary examination of the efficacy of the \emph{miniPXI} in the context of early-stage game development was undertaken in a recent study~\cite{haider2023}. The study demonstrated the effectiveness of the \emph{miniPXI} in distinguishing PX across several prototype iterations. Interestingly, the survey also included a slightly tailored version of \emph{Net Promoter Score} (\emph{NPS})~\cite{reichheld2003one}, a well-established market research metric that is based on a single item asking respondents to rate the likelihood that they would recommend a company to their friends. Instead of inquiring about recommending a company, the tailored version inquired about the likelihood of players recommending the game to their friends. The preliminary results suggest a potential for applying the \emph{NPS} in PX evaluations to get a general overview of player satisfaction from a game.
The \emph{NPS} also outperformed the \emph{miniPXI}'s enjoyment item, included as an umbrella item to the ten original \emph{PXI} constructs. However, the study involved a limited number of participants ($n = 16$). In summary, the consistency of single-item per construct measures, such as the \emph{miniPXI} and a single-item measure like the \emph{NPS}, in a scenario involving a repeated evaluation of PX across different games and including a comparison with multi-item measures, is not yet adequately studied. 

This study aims to close this gap by examining the test-retest reliability of the \emph{miniPXI} and how comparable it is to already established multi-item measures for PX evaluation. Additionally, we assess the test-retest reliability of the slightly adapted version of the \emph{NPS}~\cite{reichheld2003one}, compared to the \textsc{Enjoyment} item of the \emph{miniPXI}. As such, we seek to address the following research questions:

\begin{description}
    \item[RQ1:] How does the \emph{miniPXI} perform with respect to test-retest reliability?
    \item[RQ2:] How does the \emph{miniPXI} perform in test-retest reliability, when compared to other popular measures that use multiple items per construct?
    \item[RQ3:] How does the overall enjoyment item of the \emph{miniPXI} compare in test-retest reliability to the \emph{NPS}?
\end{description}

\noindent To this end, our study explored the test-retest reliability of the \emph{miniPXI} and established multi-item surveys, as well as the adapted \emph{NPS} across four different games, representing different genres and with 100 participants. 

This paper makes four key contributions. First, it shows that while single-item variants to measure PX, i.e., the \emph{miniPXI} demonstrate various degrees of repeatability across different constructs and genres, multi-item measures are shown to be more stable. Second, the paper systematically compares the test-retest reliability of different established PX measures when conducting repeated measurements of PX. Therefore, it sets a benchmark for the test-retest reliability to be expected across different games and PX dimensions. Third, it also contributes to a theoretical understanding of the inherent (in)repeatability of certain dimensions of PX. Finally, it highlights the viability of the adapted \emph{NPS} as a single-item measure for PX evaluation. In summary, our findings underscore the need for a more refined understanding of PX dimensions, particularly utilizing single-item per construct measures for repeated evaluation in future GUR endeavors.

\section{Related Work}
    \label{sec:relatedwork}
    
In the following, we discuss the effectiveness of single-item measures in capturing the diverse facets of PX, followed by detailing the specific aspects of the \emph{miniPXI}---a single-item per construct version of the \emph{PXI}---specifically developed to evaluate PX in various gaming contexts. Additionally, we look into the approaches used to validate single-item measures, emphasizing the intricacies and difficulties in establishing the test-retest reliability of such measures.

\subsection{Single-Item Measures for Evaluating PX}
In the field of GUR, a variety of techniques for measuring PX exists~\cite{drachen_introduction_2018}. These methods often encompass game metrics~\cite{Guenter2019}, biometric measurements~\cite{pejman_biometrics_2012}, artificial intelligence algorithms~\cite{makantasis_AI_2019}, and playability heuristics~\cite{Desurvire_heuristics_2009}. While these approaches provide valuable insights, methods that allow players to self-report on specific aspects of the experience, continue to be one of the most effective methods for assessing PX comprehensively~\cite{Abeele2020}. Such `introspective' measures allow players to provide `subjective' feedback and insights into their personal experiences, thoughts, and emotions while playing a game. Therefore, in the past decades, a number of validated measures have been created, among others, the \emph{Game Engagement Questionnaire} (\emph{GEQ})~\cite{BROCKMYER_GEQ_2009} with 19 items for four dimensions, the \emph{Player Experience of Need Satisfaction Questionnaire} (\emph{PENS})~\cite{Ryan_PENS_2006} with 21 items for five dimensions, the \emph{Ubisoft Perceived Experience Questionnaire} (\emph{UPEQ})~\cite{azadvar_upeq_2018} with 21 items for three dimensions, the \emph{Player Experience Inventory} (\emph{PXI})~\cite{Abeele2020} with 33 items for eleven dimensions, and related user experience (UX) measures such as the \emph{AttrakDiff}~\cite{Hassenzahl2003}, with 28 items for four dimensions. Consequently, these measures---involving between 19 and 33 items, measuring between three and eleven dimensions---are sometimes described as lengthy by industry partners~\cite{VERKUYL202216} and resource demanding~\cite{kaisa2008}.

In contrast, single-item per-construct measures, also known as single-item measures, might be advantageous for practical user research situations. They allow for quick and efficient data collection and integrate well into iterative evaluations with tight schedules and budgets~\cite{haider_minipxi_2022}. These measures have also been reported as offering greater face validity, allowing scores to be more easily interpreted and compared across implementations~\cite{Dolbier_2004}. They also exhibit less variation in adaptation across populations and contexts, minimize missing or invalid responses, and mitigate participant fatigue~\cite{Drolet2001,Wanous1997}.  
 
Hence, such short measures are particularly advantageous for studies with repeated measurements---as in the context of iterative game development, in the case of autonomous questionnaire completion (e.g., online or mobile studies), or where PX is only one of many aspects to be evaluated (e.g., in the context of serious games, where researchers may also want to assess persuasiveness or pedagogic qualities which increases study times). 

However, there are also limitations to single-item measures. In general, due to their design, single-item per construct measures only assess one aspect of a complex concept. As a result, they are more likely to be affected by random errors and may be less effective in capturing individual variations or changes over time~\cite{DeVellis2017}. In turn, single-item measures may encounter difficulties in fully capturing the full nature of more complex psychological constructs~\cite{Fisher2016DevelopingAI} such as PX\cite{josefPX2016}. This can be particularly problematic when assessing and differentiating among ambiguous constructs~\cite{Loo2002} such as flow~\cite{csikszentmihalyi_flow:_1990,susan1996} or immersion~\cite{ermi2005fundamental,brown2004}. In addition, the absence of an internal reliability assessment and the inability to distinguish between explained and unexplained variances restricts their use for more elaborate statistical modeling~\cite{nunnally_1978,Dolbier_2004}. 

\subsection{\emph{miniPXI} as a Single-Item per Construct Variant of the \emph{PXI} Questionnaire}
The \emph{PXI} is a validated measure that measures 11 constructs, utilizing three items per construct~\cite{Abeele2020}. Five constructs sit at the level of \textsc{Functional Consequences}, focusing on the immediate, tangible outcomes resulting from game design choices, including \textsc{Audiovisual Appeal}, \textsc{Progress Feedback}, \textsc{Clarity of Goals}, \textsc{Challenge}, and \textsc{Ease of Control}. 
Five constructs sit at the level of the \textsc{Psychosocial Consequences}, exploring emotional experiences as second-order responses to game design choices and encompass the constructs of \textsc{Immersion}, \textsc{Mastery}, \textsc{Meaning}, \textsc{Curiosity}, and \textsc{Autonomy}. Finally, participants’ overall \textsc{Enjoyment} of the game is measured as an umbrella single item. Recently, a short variant of the PXI was proposed: the \emph{miniPXI}~\cite{haider_minipxi_2022}, which reduces the \emph{PXI} to one item per construct, resulting in an 11-item measure.

A validation study of the \emph{miniPXI} provided nuanced results; reliability estimates for the \emph{PXI} constructs varied, and the authors could only confirm the validity for nine out of 11 constructs~\cite{haider_minipxi_2022}. The reliability and validity results indicated that the short version did not perform at the same level as the full variant. Additionally, a preliminary study assessed the effectiveness of the \emph{miniPXI} within the framework of early-stage game development~\cite{haider2023}. The \emph{miniPXI}'s ability to distinguish PX among prototype iterations was confirmed in this study. Findings showed that when differences in nearly all PX dimensions were considered during iterations, the \emph{miniPXI} demonstrated notable validity at the level of individual games. Interestingly, this study also included alternative metrics, among them the \emph{NPS}. This metric, while not measuring PX directly but rather using the intention of recommending the game, was found promising as a potential proxy for single-item measurement of PX to capture the overall players' satisfaction. 
Hence, the study demonstrated that the \emph{miniPXI} can assess iterations of game prototypes, beginning at the earliest stages of game development. Nevertheless, as a preliminary study, the number of players was modest, only including 16 players. 

\subsection{Internal vs. Test-Retest Reliability of Single-Item Measures}
Internal and test-retest reliabilities are important indicators of the effectiveness of single-item measures. Internal reliability is the consistency of responses within a measurement tool, which is often examined using reliability coefficients, i.e., Cronbach's $\alpha$~\cite{Cronbach1951} and McDonald's $\omega$~\cite{mcdonald}. Test-retest reliability, on the other hand, measures the consistency of scores over time and is often assessed by calculating the Intraclass Correlation Coefficient (ICC). Based on the guidelines provided by~\citet{KOO2016155}, on the 95\% confidence interval (CI) of the ICC estimate, values less than 0.5, between 0.5 and 0.75, between 0.75 and 0.9, and greater than 0.90 are indicative of poor, moderate, good, and excellent test-retest reliability, respectively. \citet{reinout2016} examined the factors influencing single-item internal reliability, test-retest reliability, and self-other agreement of single-item measures for personality assessment. One of their main arguments was to prioritize test-retest reliability over single-item internal reliability when assessing the effectiveness of a single-item measure. Test-retest reliability is a more valid measure of personality trait consistency than internal reliability, as it is less susceptible to the influence of \textit{evaluativeness}, observability, item domain, and item position~\cite{reinout2016}. Similarly, \citet{robert2011} emphasized that while the internal reliability of measures can serve as a valuable tool for assessing data quality, it only represents the coherence (or redundancy) of the components of a measure. It is also conceptually independent of test-retest reliability, which instead reflects the degree to which similar scores are achieved when the measure is administered on several occasions separated by a relatively short interval~\cite{robert2011}.

Regarding the time span, there is no agreement on the optimal duration for measuring test-retest reliability; this largely remains domain dependent. In the Human-Computer Interaction (HCI) domain, \citet{zniak2021} used a 14-day time span for their studies, although other domains, like personality factors and sleeping quality studies, used different lengths for repeated measurements ~\cite{Cattell2001,WATSON2004319,BACKHAUS2002737}. Interestingly, in the health domain, \citet{Marx2003-ph} found no significant distinction between using a two-day gap and a two-week delay between the repeated measures. Unfortunately, to the best of the authors' knowledge, no similar studies were found within the domain of user or player experiences that suggest appropriate time spans for investigating test-retest reliability.
\par

\noindent In conclusion, the question remains whether the \emph{miniPXI} is stable over time in capturing the diverse constructs of PX. Therefore, this study aims to assess the test-retest reliability of the \emph{miniPXI} over a duration of three weeks.

\section{Methodology}
    \label{sec:methodology}
    
To assess the test-retest reliability of the \emph{miniPXI}, we utilized a repeated measures setup over a time span of three weeks to capture variations in PX over time. Participants first took part in gaming sessions involving four distinct games of different genres, enabling us to evaluate the performance of the \emph{miniPXI} across genres. Additionally, established multi-item measures were used in conjunction with the \emph{miniPXI}, to compare the repeatability of the different measures. Three weeks later, participants were again asked to play the same game and fill out the same measures assigned to them during the first week. This allowed us to evaluate the test-retest reliability of the \emph{miniPXI} and compare it to other measurements. Ethical approval was provided by the University Ethics Board for the study under number [Anonymized]. 
    
\subsection{Participants}
Participants were recruited via \emph{Prolific}~\cite{prolific}, a sufficiently reliable platform commonly utilized for research studies~\cite{douglas2023} that can be utilized in longitudinal studies~\cite{kothe2019retention}. The specified recruitment criteria required participants to be fluent in English as all provided instructions were in English, and to be aged between 18 and 99 years. We also set a high approval rate threshold, only including those with an approval rate between 98\% and 100\% on \emph{Prolific}. 
The types of games included computer games, console games, handheld console games, free-to-play mobile games, premium mobile games (pay to download), esports games, virtual reality games, and online casino games. Additionally, participants needed to list video games as one of their hobbies. Participants were compensated at the rate of 11€ per hour for their participation in both sessions. From the initial sample of 104 responses, four were dropped based on suspicious answer patterns (graphic patterns or limited diversity in answers)~\cite{Meade2012}, leaving a final sample of 100 participants for further analysis. 

The age of the participants ranged from 20 to 55 years, with a mean of 28.74 years ($SD = 6.98$) and a median age of 26.50 years. With respect to gender, 47\% ($n=47$) were men, 46\% ($n=46$) women, 6\% ($n=6$) identified as non-binary, and 1\% ($n=1$) preferred to self describe. Regarding the hours per week spent by participants playing games, 1\% ($n=1$) indicated they spend 1 to 2 hours, 19.0\% ($n=19$) spend 2 to 5 hours, 30.0\% ($n=30$) spend 5 to 10 hours, 22.0\% ($n=22$) spend 10 to 20 hours, and 28.0\% ($n=28$) spend more than 20 hours per week. Participants also self-rated their expertise levels in gaming on a 7-point scale from \textit{novice (1)} to \textit{expert (7)}. The majority of the participants (84\%, $n=84$) considered themselves to be proficient to expert gamers, i.e., rating $\geq 5$, followed by 13\% ($n=13$) at competent level with a rating of 4, and a smaller proportion of participants (3\%, $n=3$) identified themselves as novices to advanced beginners with a rating $\leq 3$. Participants were distributed geographically as follows: 69\% ($n = 69$) from Europe , 14\% ($n = 14$) from North America, 9.0\% ($n = 9$) from Africa, 5\% ($n = 5$) from South America, and 3\% ($n = 3$) from Asia.

\subsection{Measures}
    \label{subsec:measures}

In addition to the \emph{miniPXI} and the \emph{PXI}, commonly used measures for evaluating PX were leveraged, including the \emph{Player Experience of Need Satisfaction} questionnaire (\emph{PENS})~\cite{Ryan_PENS_2006}, the \emph{Game Engagement Questionnaire} (\emph{GEQ})~\cite{BROCKMYER_GEQ_2009}, and a related UX measure, the \emph{AttrakDiff}~\cite{Hassenzahl2003}. The original work on the \emph{PXI}~\cite{Abeele2020} served as the primary criteria for including the \emph{PENS} and \emph{AttrakDiff} as additional measures. Moreover, the \emph{GEQ} was included, as it is among the most widely used and well-studied measures in player experience research ~\cite{denisova_convergence_2016, JOHNSON201838}. 

\begin{itemize}
    \item The \emph{Player Experience of Need Satisfaction} questionnaire is a 21-item measure, primarily based on self-determination theory, that aims to assess psychological needs fulfillment within gaming contexts. Utilizing a 7-point Likert scale ranging from \textit{Do Not Agree (1)} to \textit{Strongly Agree (7)}, it employs items focused on dimensions such as autonomy, competence, relatedness, and immersion. 
    \item The \emph{Game Engagement Questionnaire} is a validated measure designed to evaluate the level of engagement experienced by players during gaming sessions. Comprising 19 items, each rated on a 5-point Likert scale, ranging from \textit{Not at all (1)} to \textit{Sort of (3)} to \textit{Yes (5)}, the \emph{GEQ} assesses multiple dimensions of game engagement, with items evaluating (cognitive) absorption, immersion, flow, and presence. 
    \item The \emph{AttrakDiff}, comprised of 28 items rated on a 7-point semantic differential scale, is a measure designed to capture both the pragmatic qualities (e.g., usability, efficiency) and hedonic aspects (e.g., stimulation, aesthetics) of the UX.
    \item The \emph{Net Promoter Score} (\emph{NPS})~\cite{reichheld2003one} is a metric used to assess consumer satisfaction and loyalty. In the current study, a slightly adapted version of the \emph{NPS} to the context of gaming is used to evaluate player satisfaction and the probability of recommending a particular game to others. The survey consists of a single item: \textit{How likely are you to recommend this game to a friend or colleague?}, assessed on a scale from 0 to 10. The ratings of the responses fall into three categories: promoters (scoring 9-10), passives (scoring 7-8), and detractors (scoring 0-6). The \emph{NPS} calculation can possibly provide valuable information regarding player satisfaction and loyalty by subtracting the proportion of detractors from the proportion of promoters. 
\end{itemize}

\subsection{Games}
\begin{figure*}
    \centering
    \hfill
    \begin{minipage}[t]{0.48\linewidth}
        \centering
        \includegraphics[width=1.0\linewidth]{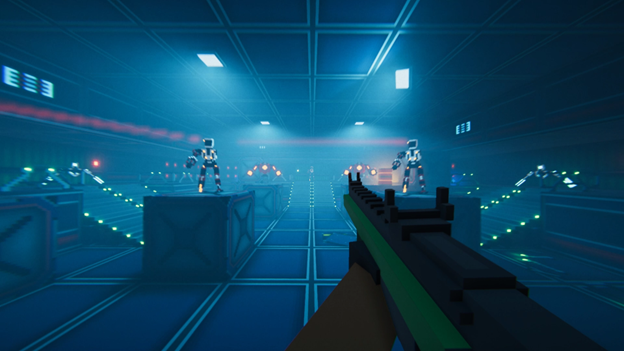}
        (a) \emph{Station Saturn} (First Person Shooter)
    \end{minipage}
    \hfill
    \begin{minipage}[t]{0.48\linewidth}
        \centering
        \includegraphics[width=1.0\linewidth]{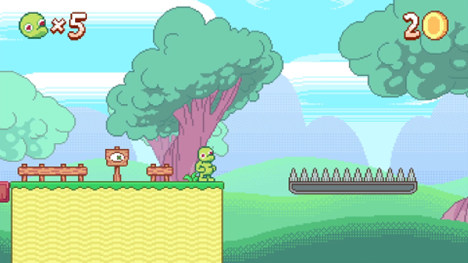}
        (b) \emph{Little Runmo} (Platformer) 
    \end{minipage}
    \hfill
    \vskip5pt
    \hfill
    \begin{minipage}[t]{0.48\linewidth}
        \centering
        \includegraphics[width=1.0\linewidth]{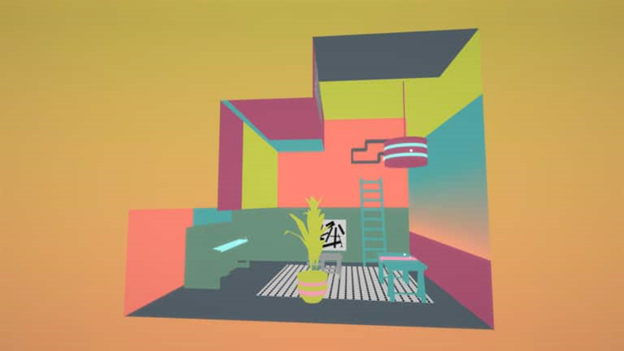}
        (c) \emph{Empty} (Puzzle)
    \end{minipage}
    \hfill
    \begin{minipage}[t]{0.48\linewidth}
        \centering
        \includegraphics[width=1.0\linewidth]{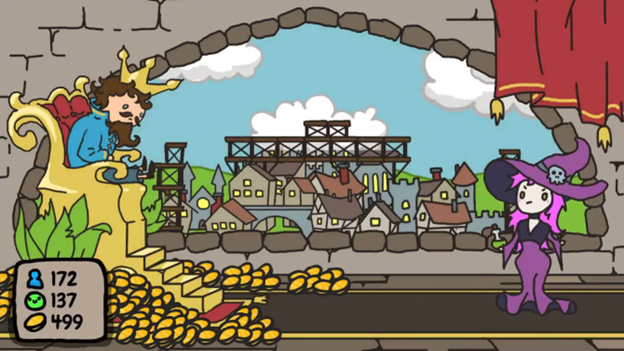}
        (d) \emph{Sort the Court} (Simulation)
    \end{minipage}
    \hfill
    \caption{Screenshots of the four games used in the study. Each game represents a different genre (noted in brackets).}
    \Description{Screenshots of the four games included in the study.}
    \label{fig:games}
\end{figure*}

The games were chosen based on specific criteria that they should be fully developed, free to play, browser based to facilitate remote participation, playable in short duration (15-20 minutes), and use familiar controls to ease the onboarding process and minimize learning curves.
Additionally, the focus was on selecting games that represent different genres. Based on that, first, eight games from \emph{itch.io}~\cite{itchio}---a website for browser-based indie games---were shortlisted with input from three GUR experts. Next, in agreement with all co-authors and following the guidelines for selecting games for research projects~\cite{tyack2018}, four games (see Figure~\ref{fig:games}) were selected, each representing a different genre. 

\begin{description}[labelindent=2.0\parindent,leftmargin=2.0\parindent,itemsep=0.5\baselineskip]
    \item[Station Saturn]\cite{saturn} (Figure~\ref{fig:games}a) is a fast-paced, infinite first-person shooter game, set within a science fiction theme. The game pits players against robots within enclosed rooms. Killing all enemies in a room will allow the player to advance to the next. It uses established controls common to this genre, that is, WASD keys for navigation and the mouse to look around and shoot. 
    
    \item[Little Runmo]\cite{runmo} (Figure~\ref{fig:games}b) is a 2D platformer based on the animated \emph{YouTube} video of the same name~\cite{youtubeLittleRunmo}. The storyline follows the journey of a fictional video game character drawing inspiration from the original video and the classic \emph{Super Mario} series, that is, the player needs to collect coins, avoid enemies or jump on them to defeat them. In addition, the character sticks to walls on which it slowly glides down.
    
    \item[Empty]\cite{empty} (Figure~\ref{fig:games}c) is a minimalist puzzle game that is based on innovative puzzle mechanics. The player's primary goal is manipulating the room's rotation via the mouse to align objects with the same color to remove them visually. Due to the objects occluding each other, the player also needs to find out the correct order to dissolve them. 
    
    \item[Sort the Court]\cite{court} (Figure~\ref{fig:games}d) is a simulation game that invites players to rule a kingdom by focusing on decision-making and strategic thinking. Players assume the role of a monarch, balancing choices to manage resources, expand the kingdom, and ensure the well-being of its citizens. The game's core mechanic revolves around the player responding with 'yes' or 'no' to various requests of citizens, which, in turn, will influence the kingdom's prosperity and growth. 
\end{description}

\subsection{Procedure}
\begin{figure*}
    \centering
    \includegraphics[width=0.9\linewidth]{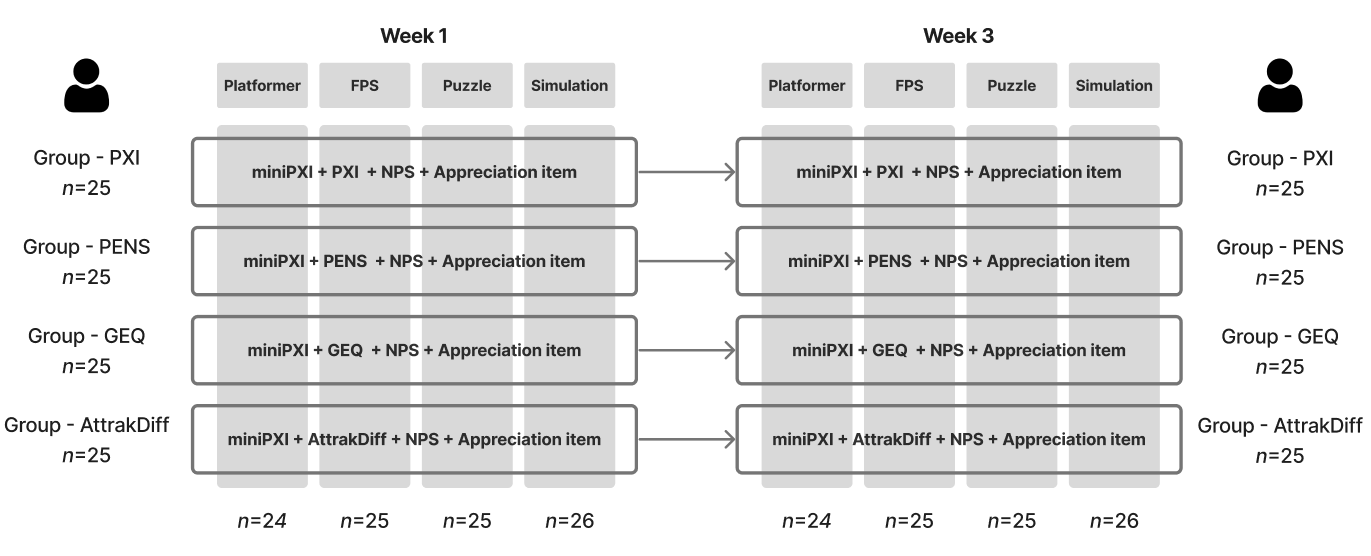}
    \caption{The experimental setup for the test-retest study with 100 participants, two measurements (3-week gap), four established measures across four games.}
    \Description{Flowchart illustrating a test-retest study with 100 participants, two measurements (3-week gap), and four established measures across four games.}
    \label{Fig.flow}
\end{figure*}

The study employed a test-retest method with a three-week span to assess the reliability and repeatability of the \emph{miniPXI} in capturing PX over multiple gaming weeks, see Figure~\ref{Fig.flow}. \emph{Qualtrics}~\cite{qualtrics} was used to create and administer the online survey. During the first week, participants were asked to provide informed consent, demographic information, and insights into their gaming habits. Then, each participant was randomly assigned, counterbalanced, to one of the four games and asked to spend 8 to 10 minutes playing the assigned game. Afterward, they were asked to rate their experience using the \emph{miniPXI} and one additional established PX measures included in the study (\emph{PXI}, \emph{PENS}, \emph{AttrakDiff}, or \emph{GEQ}) (cf. Section~\ref{subsec:measures}), to which they were also assigned randomly and counterbalanced. Only one additional multi-item measure per group was included to ensure a reasonable survey duration and to prevent participant fatigue. 

Lastly, all participants were asked to respond to the adapted \emph{NPS} item, worded as \textit{"How likely are you to recommend this game to a friend or colleague?"}. In addition, we added an 'appreciation' item, worded as \textit{''Overall, I appreciate the game''}. This last item, similar to \emph{NPS}, does not directly measure PX but rather asks for the overall appreciation and was added to compare performance to the NPS item. In short, each participant was assigned to one game and a group of measures (\emph{miniPXI} + (\emph{PXI} | \emph{PENS} | \emph{GEQ} | \emph{AttrakDiff}) + \emph{NPS} + Appreciation Item). We will use the names of the games and of the multi-item measures to refer to the respective groups.  

After a three-week break, the participants were invited again to engage with the same game they were assigned to in the first week for a similar duration. Subsequently, they were asked to reassess their experience, using the exact same measures they used in the first week (\emph{miniPXI} + multi-item measure + \emph{NPS} + Appreciation item). 

\subsection{Analysis}
\emph{Jamovi} v2.3.28~\cite{jamovi} was used for most of the statistical analyses. Additionally, \emph{R} v4.3.0~\cite{coreteamr} was used for data pre-processing, which was performed based on the procedure recommended by Gaskin~\cite{gaskin2012data}. 

\subsubsection{Internal Consistency}
In order to verify the reliabilities of multi-item measures used, the internal consistencies of the multi-item per construct measures used in this study, namely the \emph{PXI}, \emph{PENS}, \emph{GEQ}, and \emph{AttrakDiff}, were assessed by calculating Cronbach's $\alpha$~\cite{Cronbach1951} and McDonald's $\omega$~\cite{mcdonald} for each of the constructs and for both weeks, i.e. week 1 and week 3. Both coefficients are sensitive to the coding direction for Likert scale data. As some of the items from the \emph{PENS}, \emph{GEQ}, and \emph{AttrakDiff} were inverted, these questions were reversed before computing the coefficients. Cronbach's $\alpha$ values obtained were assessed against established criteria for satisfactory internal consistency: according to \citet{field2013discovering}, a $\alpha$ coefficient greater than $0.7$ is commonly considered to signify satisfactory internal consistency in a given scale. 

\subsubsection{Test-retest Reliability}
There are multiple ways in which the test-retest reliability of an instrument can be assessed, as suggested by \citet{boateng2018}. Based on the guidelines provided by \citet{KOO2016155} and recommendations of \citet{McGraw1996}, the intra-class correlation coefficient (ICC) from a 2-way mixed-effects model with the absolute agreement was used to investigate the relationship between the two measurement moments.\footnote{Different measures besides the ICC can be used to investigate test-retest reliability. Given that the ICC is currently promoted as the most rigorous metric, we provide this metric in the results. However, other additional measures, such as Pearson's $r$ or NRMSE, can be found in the supplemental materials.} Based on the guidelines provided by \citet{KOO2016155}, on the 95\% confident interval of the ICC estimate, values less than 0.5, between 0.5 and 0.75, between 0.75 and 0.9, and greater than 0.90 are respectively indicative of poor, moderate, good, and excellent test-retest reliability. For further granularity, we have split the poor category into two halves, i.e., less than 0.25 and between 0.25 and 0.5, indicative of very poor and poor, respectively.  

\begin{itemize}[itemsep=0.5\baselineskip]
    \item For \textbf{RQ1}, the test-retest reliability of the \emph{miniPXI} was examined in three different axes. First, ICC values of each of the \emph{miniPXI} constructs using the entire dataset were computed, which included all 100 responses, to see overall repeatability with a comparatively bigger sample. Next, ICC values for every construct across each of the four games were computed individually (vertical slices in Figure~\ref{Fig.flow}), to assess the repeatability of the \emph{miniPXI} when a single game is evaluated over time. Finally, we computed ICC values for each of the groups based on the accompanying measures and across multiple genres (horizontal slices in Figure~\ref{Fig.flow}), i.e., to assess the repeatability of the \emph{miniPXI} in situations where varying genres are involved. 
    
    \item For \textbf{RQ2}, the ICC values were computed for each of the constructs in multi-item measures, i.e., the \emph{PXI}, \emph{PENS}, \emph{GEQ}, and \emph{AttrakDiff}, to assess their test-retest reliability. The values were then compared with values of the \emph{miniPXI}. 
    
    \item For \textbf{RQ3}, similar to RQ1, the ICC values were computed and assessed for the \emph{NPS}, the 'appreciation' item, and the overall \textsc{Enjoyment} item of the \emph{miniPXI}, in three different axes described above (RQ1). 
\end{itemize}

\section{Results}
    \label{sec:results}

Below, we discuss the outcomes in detail with respect to each of the research questions. 

\begin{table}
\centering
\caption{Mean and standard deviations ($SD$) across Week 1 and 3, Intra-class correlation coefficient (ICC) interpretation in terms of test-retest reliability (
$0.25 \leq ICC < 0.5$: poor = \textcolor{cPoor}{\rule{2.5ex}{1.5ex}}; $0.5 \leq ICC < 0.75$: moderate = \textcolor{cModerate}{\rule{2.5ex}{1.5ex}}; $0.75 \leq ICC < 0.9$: good = \textcolor{cGood}{\rule{2.5ex}{1.5ex}}
), 95\% confidence interval (CI) lower (L) and upper (U) bound, Internal Consistency (IC) (Cronbach’s $\alpha$, and McDonald’s $\omega$) for multi-item measures across Week 1 and 3.}
\vskip-9pt
\label{table:all_questionnaires}

\begin{tabularx}{1.0\linewidth}{ll@{\hskip1pt} c@{\hskip4pt}c@{\hskip6pt} c@{\hskip6pt} c@{\hskip6pt}c@{\hskip6pt} c@{\hskip6pt}c@{\hskip6pt} c@{\hskip6pt}c@{\hskip3pt}}
    \toprule
     &  &  &  &    &  &  & \multicolumn{4}{c}{\hskip-5pt\textbf{IC}} \\

     \cmidrule(lr){8-11}
     
     &  & \multicolumn{2}{c}{\hskip-15pt\textbf{Mean ($SD$)}}   &  & \multicolumn{2}{c}{\hskip-15pt\textbf{95\% CI}} & \multicolumn{2}{c}{\hskip-15pt\textbf{Week 1}} & \multicolumn{2}{c}{\hskip-5pt\textbf{Week 3}} \\

     \cmidrule(lr){3-4} \cmidrule(lr){6-7} \cmidrule(lr){8-9} \cmidrule(lr){10-11}
     
     & \textbf{Construct} & \textbf{Week 1} & \textbf{Week 3} & \textbf{ICC} & \textbf{L} & \textbf{U} & $\bm{\alpha}$ & $\bm{\omega}$ & $\bm{\alpha}$ & $\bm{\omega}$ \\
     \midrule
     \parbox[t]{2mm}{\multirow{11}{*}{\rotatebox[origin=c]{90}{\textbf{miniPXI}}}} 
     & Audiovisual Appeal \textit{(1 item)}  & 5.65 (1.35) & 5.69 (1.35) &  \icc{.630} & 0.496 & 0.735 & -    & -    & -    & -    \\
     & Challenge \textit{(1 item)}           & 4.80 (1.51) & 5.12 (1.53) &  \icc{.422} & 0.249 & 0.570 & -    & -    & -    & -    \\
     & Ease of Control \textit{(1 item)}     & 6.01 (1.38) & 6.06 (1.16) &  \icc{.479} & 0.312 & 0.617 & -    & -    & -    & -    \\
     & Clarity of Goals \textit{(1 item)}    & 5.72 (1.61) & 5.77 (1.51) &  \icc{.624} & 0.488 & 0.731 & -    & -    & -    & -    \\
     & Progress Feedback \textit{(1 item)}   & 5.11 (1.61) & 5.00 (1.75) &  \icc{.651} & 0.522 & 0.751 & -    & -    & -    & -    \\
     & Autonomy \textit{(1 item) }          & 5.32 (1.50) & 5.12 (1.47) &  \icc{.505} & 0.345 & 0.637 & -    & -    & -    & -    \\
     & Curiosity \textit{(1 item)  }        & 5.64 (1.24) & 5.69 (1.40) &  \icc{.547} & 0.393 & 0.671 & -    & -    & -    & -    \\
     & Immersion \textit{(1 item)   }       & 6.19 (1.10) & 6.07 (1.17) &  \icc{.365} & 0.183 & 0.523 & -    & -    & -    & -    \\
     & Mastery \textit{(1 item)    }        & 5.43 (1.43) & 5.43 (1.22) &  \icc{.447} & 0.274 & 0.591 & -    & -    & -    & -    \\
     & Meaning \textit{(1 item)  }          & 4.72 (1.49) & 4.72 (1.53) &  \icc{.644} & 0.513 & 0.746 & -    & -    & -    & -    \\
     & Enjoyment \textit{(1 item)  }        & 5.88 (1.24) & 5.88 (1.26) &  \icc{.704} & 0.590 & 0.791 & -    & -    & -    & -    \\
     \midrule
     \multicolumn{2}{l}{\textbf{NPS} \textit{(1 item)}} & 6.21 (2.65) & 6.29 (2.74) & \icc{.791} & 0.704 & 0.854 & -    & -    & -    & -    \\
     \midrule
     \multicolumn{2}{l}{\textbf{Appreciation} \textit{(1 item)}} & 5.70 (1.28) & 5.73 (1.25) & \icc{.795} & 0.710 & 0.857 & -    & -    & -    & -    \\
     \midrule
     \parbox[t]{2mm}{\multirow{11}{*}{\rotatebox[origin=c]{90}{\textbf{PXI}}}}
     & Audiovisual Appeal (\textit{3 items) }  & 5.92 (1.00) & 5.92 (0.90) & \icc{.668} & 0.374 & 0.840 & 0.95 & 0.96 & 0.94 & 0.94 \\
     & Challenge \textit{(3 items) }           & 5.12 (1.37) & 5.03 (1.23) & \icc{.820} & 0.634 & 0.916 & 0.88 & 0.88 & 0.8  & 0.81 \\
     & Ease of Control \textit{(3 items)   }   & 6.08 (1.00) & 6.20 (0.71) & \icc{.643} & 0.341 & 0.825 & 0.90 & 0.91 & 0.78 & 0.81 \\
     & Clarity of Goals \textit{(3 items) }    & 5.88 (1.03) & 5.72 (1.01) & \icc{.856} & 0.701 & 0.934 & 0.89 & 0.90 & 0.89 & 0.89 \\
     & Progress Feedback \textit{(3 items) }   & 5.41 (1.26) & 5.53 (1.48) & \icc{.664} & 0.370 & 0.837 & 0.88 & 0.89 & 0.90 & 0.90 \\
     & Autonomy \textit{(3 items) }          & 5.05 (1.37) & 5.04 (1.32) & \icc{.657} & 0.356 & 0.833 & 0.91 & 0.91 & 0.91 & 0.91 \\
     & Curiosity (\textit{3 items)  }         & 6.04 (0.67) & 5.87 (1.07) & \icc{.495} & 0.137 & 0.740 & 0.90 & 0.91 & 0.94 & 0.94 \\
     & Immersion \textit{(3 items)  }        & 5.77 (0.87) & 5.65 (1.13) & \icc{.740} & 0.496 & 0.876 & 0.71 & 0.79 & 0.81 & 0.85 \\
     & Mastery \textit{(3 items)  }           & 5.53 (1.02) & 5.56 (1.00) & \icc{.566} & 0.221 & 0.784 & 0.79 & 0.79 & 0.72 & 0.76 \\
     & Meaning \textit{(3 items) }            & 4.99 (1.35) & 5.04 (1.30) & \icc{.881} & 0.749 & 0.946 & 0.96 & 0.96 & 0.92 & 0.92 \\
     & Enjoyment \textit{(3 items)  }         & 5.96 (0.85) & 6.07 (0.67) & \icc{.649} & 0.351 & 0.828 & 0.94 & 0.94 & 0.86 & 0.88 \\
     \midrule
     \parbox[t]{2mm}{\multirow{5}{*}{\rotatebox[origin=c]{90}{\textbf{PENS}}}}
     & Competence \textit{(3 items) }          & 5.57 (1.13) & 5.48 (1.24) & \icc{.470} & 0.093 & 0.727 & 0.85 & 0.86 & 0.93 & 0.94 \\
     & Autonomy \textit{(3 items) }           & 4.20 (1.45) & 4.37 (1.70) & \icc{.695} & 0.422 & 0.853 & 0.80 & 0.82 & 0.89 & 0.89 \\
     & Relatedness  \textit{(3 items)  }      & 2.88 (1.53) & 2.91 (1.65) & \icc{.689} & 0.407 & 0.851 & 0.77 & 0.78 & 0.81 & 0.85 \\
     & Presence/Immersion \textit{(8 items)}  & 3.37 (1.32) & 3.36 (1.39) & \icc{.734} & 0.480 & 0.874 & 0.88 & 0.89 & 0.92 & 0.92 \\
     & Intuitive Controls \textit{(3 items) } & 6.17 (0.77) & 6.03 (1.18) & \icc{.589} & 0.262 & 0.795 & 0.66 & 0.74 & 0.87 & 0.91 \\
     \midrule
     \parbox[t]{2mm}{\multirow{4}{*}{\rotatebox[origin=c]{90}{\textbf{GEQ}}}}
     & Absorption \textit{(5 items)  }        & 2.32 (0.89) & 2.33 (0.73) & \icc{.848} & 0.684 & 0.930 & 0.81 & 0.82 & 0.77 & 0.77 \\
     & Flow  \textit{(9 items)  }             & 2.60 (0.73) & 2.66 (0.74) & \icc{.852} & 0.695 & 0.932 & 0.79 & 0.80 & 0.82 & 0.83 \\
     & Presence \textit{(4 items)}           & 2.98 (0.96) & 2.86 (0.94) & \icc{.765} & 0.540 & 0.889 & 0.66 & 0.70 & 0.76 & 0.79 \\
     & Immersion \textit{(1 item)  }       & 3.44 (1.19) & 3.52 (1.23) & \icc{.753} & 0.514 & 0.883 & -    & -    & -    & -    \\
     \midrule
     \parbox[t]{2mm}{\multirow{4}{*}{\rotatebox[origin=c]{90}{\textbf{AttrakDiff}}}}
     & Attractiveness  \textit{ (7 items) }    & 5.44 (1.01) & 5.39 (1.00) & \icc{.750} & 0.508 & 0.882 & 0.91 & 0.91 & 0.91 & 0.92 \\
     & HQ - Identification \textit{(7 items) } & 4.45 (0.79) & 4.39 (0.98) & \icc{.811} & 0.617 & 0.912 & 0.73 & 0.74 & 0.82 & 0.83 \\
     & HQ - Stimulation \textit{(7 items) }    & 4.50 (0.98) & 4.53 (1.05) & \icc{.853} & 0.694 & 0.932 & 0.82 & 0.84 & 0.84 & 0.85 \\
     & Pragmatic Quality \textit{(7 items)  } & 4.94 (0.87) & 4.80 (0.70) & \icc{.715} & 0.457 & 0.862 & 0.71 & 0.79 & 0.57 & 0.59 \\
     \bottomrule
\end{tabularx}

\begin{flushleft}
\footnotesize{$n=100$ for \emph{miniPXI}, \emph{NPS}, and appreciation; $n=25$ in all other cases, HQ = hedonic quality.}
\end{flushleft}
\end{table}

\afterpage{%
\begin{landscape}
\begin{table}
\fontsize{7.9}{9.8}\selectfont
\centering
\caption{Mean and standard deviations ($SD$) across Week 1 and 3, intra-class correlation coefficient (ICC) interpretation in terms of test-retest reliability ($ICC < 0.25$: very poor = \textcolor{cVpoor}{\rule{2.5ex}{1.5ex}}; $0.25 \leq ICC < 0.5$: poor = \textcolor{cPoor}{\rule{2.5ex}{1.5ex}}; $0.5 \leq ICC < 0.75$: moderate = \textcolor{cModerate}{\rule{2.5ex}{1.5ex}}; $0.75 \leq ICC < 0.9$: good = \textcolor{cGood}{\rule{2.5ex}{1.5ex}} 
), 95\% confidence interval (CI) lower (L) and upper (U) bound, for each construct (C) across each of the four games.}
\vskip-0.5\baselineskip
\label{table:minipxi_games}
\setlength{\tabcolsep}{1.7pt}
\def\arraystretch{0.86}
\begin{tabularx}{1.0\linewidth}{l cccd{2.3}d{1.3}@{\hskip6pt} cccd{2.3}d{1.3}@{\hskip6pt} cccd{2.3}d{1.3}@{\hskip6pt} cccd{2.3}d{1.3}}
\toprule
&  \multicolumn{5}{c}{\textbf{Station Saturn (First-Person Shooter)}} & \multicolumn{5}{c}{\textbf{Little Runmo (Platformer)}} & \multicolumn{5}{c}{\textbf{Empty (Puzzle game)}} & \multicolumn{5}{c}{\textbf{Sort the Court (Simulation)}} \\[3pt]

\cmidrule(lr){2-6} \cmidrule(lr){7-11} \cmidrule(lr){12-16} \cmidrule(lr){17-21}

&  \multicolumn{2}{c}{\textbf{Mean ($SD$)}} &    & \multicolumn{2}{c}{\textbf{95\% CI}} & \multicolumn{2}{c}{\textbf{Mean ($SD$)}} &    & \multicolumn{2}{c}{\textbf{95\% CI}} & \multicolumn{2}{c}{\textbf{Mean ($SD$)}} &    & \multicolumn{2}{c}{\textbf{95\% CI}} & \multicolumn{2}{c}{\textbf{Mean ($SD$)}} &    & \multicolumn{2}{c}{\textbf{95\% CI}} \\

\cmidrule(lr){2-3} \cmidrule(lr){5-6} \cmidrule(lr){7-8} \cmidrule(lr){10-11} \cmidrule(lr){12-13} \cmidrule(lr){15-16} \cmidrule(lr){17-18} \cmidrule(lr){20-21}

\textbf{C} & \textbf{Week 1} & \textbf{Week 3} &  \textbf{ICC} & \multicolumn{1}{c}{\textbf{L}} &  \multicolumn{1}{c}{\textbf{U}} & \textbf{Week 1} & \textbf{Week 3} & \textbf{ICC} & \multicolumn{1}{c}{\textbf{L}} & \multicolumn{1}{c}{\textbf{U}} & \textbf{Week 1} & \textbf{Week 3} & \textbf{ICC} & \multicolumn{1}{c}{\textbf{L}} & \multicolumn{1}{c}{\textbf{U}} & \textbf{Week 1} & \textbf{Week 3} & \textbf{ICC} & \multicolumn{1}{c}{\textbf{L}} & \multicolumn{1}{c}{\textbf{U}} \\

\midrule
AA    & 5.38 (1.38) & 5.21 (1.41) &  \icc{.589} & 0.251  & 0.799 & 5.33 (1.69) & 5.33 (1.57) &  \icc{.840} & 0.671  & 0.926 & 6.15 (0.99) & 6.15 (1.43) &  \icc{.306} & -0.103 & 0.623 & 5.70 (1.07) & 5.81 (1.04) &  \icc{.322} & -0.075 & 0.629 \\
CH    & 5.33 (1.34) & 4.96 (1.43) &  \icc{.585} & 0.257  & 0.794 & 4.78 (1.55) & 5.11 (1.28) &  \icc{.534} & 0.191  & 0.762 & 5.19 (1.55) & 5.93 (1.14) &  \icc{.014} & -0.330 & 0.379 & 3.85 (1.38) & 4.56 (1.83) &  \icc{.318} & -0.045 & 0.615 \\
EC    & 5.92 (1.53) & 6.13 (0.80) &  \icc{.449} & 0.065  & 0.718 & 5.15 (1.83) & 5.30 (1.64) &  \icc{.549} & 0.214  & 0.771 & 6.22 (0.64) & 5.96 (1.32) &  \icc{.058} & -0.326 & 0.431 & 6.67 (0.62) & 6.67 (0.48) &  \icc{.165} & -0.246 & 0.518 \\
CG    & 5.71 (1.60) & 5.75 (1.39) &  \icc{.595} & 0.254  & 0.803 & 5.04 (1.89) & 5.26 (1.61) &  \icc{.786} & 0.573  & 0.900 & 6.30 (1.23) & 6.26 (1.38) &  \icc{.133} & -0.287 & 0.502 & 5.78 (1.53) & 5.93 (1.47) &  \icc{.739} & 0.499  & 0.874 \\
PF    & 4.88 (1.70) & 5.00 (1.84) &  \icc{.685} & 0.391  & 0.851 & 4.30 (1.61) & 4.41 (1.65) &  \icc{.791} & 0.580  & 0.903 & 5.15 (1.41) & 4.67 (2.04) &  \icc{.319} & -0.083 & 0.631 & 6.07 (1.14) & 5.78 (1.34) &  \icc{.686} & 0.418  & 0.845 \\
AUT   & 5.00 (1.74) & 4.63 (1.66) &  \icc{.767} & 0.535  & 0.892 & 5.11 (1.53) & 5.15 (1.46) &  \icc{.295} & -0.118 & 0.617 & 5.52 (1.50) & 5.07 (1.71) &  \icc{.505} & 0.155  & 0.745 & 5.59 (1.25) & 5.59 (1.12) &  \icc{.232} & -0.178 & 0.568 \\
CUR   & 5.50 (1.25) & 5.21 (1.72) &  \icc{.567} & 0.227  & 0.785 & 5.33 (1.54) & 5.44 (1.48) &  \icc{.638} & 0.332  & 0.823 & 6.04 (1.13) & 6.15 (1.35) &  \icc{.389} & 0.001  & 0.675 & 5.70 (1.10) & 5.85 (1.17) &  \icc{.295} & -0.103 & 0.610 \\
IMM   & 6.08 (1.38) & 6.25 (1.22) &  \icc{.428} & 0.034  & 0.706 & 6.04 (1.19) & 5.85 (1.29) &  \icc{.238} & -0.173 & 0.576 & 6.56 (0.58) & 6.33 (0.92) &  \icc{.456} & 0.093  & 0.715 & 6.15 (1.03) & 5.93 (1.17) &  \icc{.371} & -0.008 & 0.658 \\
MAS   & 5.50 (1.14) & 5.46 (1.38) &  \icc{.402} & -0.005 & 0.691 & 4.52 (1.53) & 4.85 (1.26) &  \icc{.596} & 0.277  & 0.798 & 6.22 (1.01) & 5.81 (1.08) &  \icc{.043} & -0.324 & 0.413 & 5.48 (1.45) & 5.63 (0.97) &  \icc{.315} & -0.080 & 0.623 \\
MEA   & 4.58 (1.50) & 4.58 (1.61) &  \icc{.365} & -0.050 & 0.665 & 4.11 (1.67) & 4.04 (1.68) &  \icc{.750} & 0.500  & 0.879 & 5.19 (1.52) & 5.26 (1.48) &  \icc{.783} & 0.565  & 0.898 & 4.67 (1.27) & 4.81 (1.36) &  \icc{.598} & 0.279  & 0.798 \\
ENJ   & 5.83 (1.27) & 5.71 (1.43) &  \icc{.597} & 0.260  & 0.804 & 5.48 (1.67) & 5.37 (1.55) &  \icc{.809} & 0.614  & 0.911 & 6.11 (1.09) & 6.11 (1.37) &  \icc{.656} & 0.369  & 0.828 & 6.00 (0.88) & 6.11 (0.85) &  \icc{.796} & 0.599  & 0.903 \\

\midrule
NPS   & 6.08 (2.62) & 5.83 (2.93) &  \icc{.816} & 0.622  & 0.916 & 4.78 (2.81) & 5.44 (2.75) &  \icc{.839} & 0.664  & 0.926 & 7.33 (2.66) & 7.41 (2.68) &  \icc{.751} & 0.518  & 0.881 & 6.44 (2.06) & 6.37 (2.48) &  \icc{.729} & 0.480 & 0.869 \\

\midrule
APR   & 5.58 (1.10) & 5.58 (1.21) &  \icc{.717} & 0.444  & 0.867 & 5.15 (1.63) & 5.22 (1.63) &  \icc{.850} & 0.689  & 0.931 & 6.15 (1.10) & 6.19 (0.89) &  \icc{.664} & 0.367  & 0.837 & 5.85 (0.95) & 5.89 (0.97) &  \icc{.771} & 0.550 & 0.890 \\
 \bottomrule
\end{tabularx}
\begin{flushleft}
\footnotesize{AA - \textsc{Audiovisual Appeal}, CH - \textsc{Challenge}, EC - \textsc{Ease of Control}, CG - \textsc{Clarity of Goals}, PF - \textsc{Progress Feedback}, AUT - \textsc{Autonomy}, CUR - \textsc{Curiosity}, IMM - \textsc{Immersion}, MAS - \textsc{Mastery}, MEA - \textsc{Meaning}, ENJ - \textsc{Enjoyment}, NPS - \emph{Net Promoter Score}, APR - Appreciation
}
\end{flushleft}
\end{table}

\begin{table}
\fontsize{7.9}{9.8}\selectfont
\centering
\caption{Mean and standard deviations ($SD$) across week 1 and 3, intra-class correlation coefficient (ICC) interpretation in terms of test-retest reliability ($ICC < 0.25$: very poor = \textcolor{cVpoor}{\rule{2.5ex}{1.5ex}}; $0.25 \leq ICC < 0.5$: poor = \textcolor{cPoor}{\rule{2.5ex}{1.5ex}}; $0.5 \leq ICC < 0.75$: moderate = \textcolor{cModerate}{\rule{2.5ex}{1.5ex}}; $0.75 \leq ICC < 0.9$: good = \textcolor{cGood}{\rule{2.5ex}{1.5ex}} 
), 95\% confidence interval (CI) lower (L) and upper (U) bounds, for each construct (C) across each of the four groups.}
\vskip-0.5\baselineskip
\label{table:minipxi_subgroups}
\setlength{\tabcolsep}{1.7pt}
\def\arraystretch{0.86}
\begin{tabularx}{1.0\linewidth}{l cccd{2.3}d{1.3}@{\hskip1pt} cccd{2.3}d{1.3}@{\hskip1pt} cccd{2.3}d{1.3}@{\hskip1pt} cccd{2.3}d{1.3}}
\toprule
 & \multicolumn{5}{c}{\textbf{Group - PXI}} & \multicolumn{5}{c}{\textbf{Group - PENS}} & \multicolumn{5}{c}{\textbf{Group  - GEQ}} & \multicolumn{5}{c}{\textbf{Group - AttrakDiff}} \\[3pt]

\cmidrule(lr){2-6} \cmidrule(lr){7-11} \cmidrule(lr){12-16} \cmidrule(lr){17-21}
 
 & \multicolumn{2}{c}{\textbf{Mean ($SD$)}} &    & \multicolumn{2}{c}{\textbf{95\% CI}} & \multicolumn{2}{c}{\textbf{Mean ($SD$)}} &    & \multicolumn{2}{c}{\textbf{95\% CI}} & \multicolumn{2}{c}{\textbf{Mean ($SD$)}} &    & \multicolumn{2}{c}{\textbf{95\% CI}} & \multicolumn{2}{c}{\textbf{Mean ($SD$)}} &    & \multicolumn{2}{c}{\textbf{95\% CI}} \\

\cmidrule(lr){2-3} \cmidrule(lr){5-6} \cmidrule(lr){7-8} \cmidrule(lr){10-11} \cmidrule(lr){12-13} \cmidrule(lr){15-16} \cmidrule(lr){17-18} \cmidrule(lr){20-21}

\textbf{C} & \textbf{Week 1} & \textbf{Week 3} & \textbf{ICC} & \multicolumn{1}{c}{\textbf{L}} & \multicolumn{1}{c}{\textbf{U}} & \textbf{Week 1} & \textbf{Week 3} &  \textbf{ICC} & \multicolumn{1}{c}{\textbf{L}} & \multicolumn{1}{c}{\textbf{U}} & \textbf{Week 1} & \textbf{Week 3} &  \textbf{ICC} & \multicolumn{1}{c}{\textbf{L}} & \multicolumn{1}{c}{\textbf{U}} & \textbf{Week 1} & \textbf{Week 3} &  \textbf{ICC} & \multicolumn{1}{c}{\textbf{L}} & \multicolumn{1}{c}{\textbf{U}} \\
\midrule
AA      & 5.88 (1.01) & 6.08 (0.95) &  \icc{.526} & 0.179 & 0.758 & 5.84 (1.31) & 5.68 (1.46) &  \icc{.730} & 0.479  & 0.871 & 5.20 (1.68) & 5.40 (1.50) &  \icc{.756} & 0.523  & 0.884 & 5.68 (1.28) & 5.60 (1.38) &  \icc{.379} & -0.022 & 0.671 \\
CH      & 5.20 (1.41) & 5.20 (1.44) &  \icc{.662} & 0.364 & 0.836 & 4.92 (1.38) & 5.08 (1.73) &  \icc{.212} & -0.205 & 0.559 & 4.08 (1.58) & 5.16 (1.68) &  \icc{.415} & 0.029  & 0.694 & 5.00 (1.50) & 5.04 (1.34) &  \icc{.484} & 0.110  & 0.736 \\
EC      & 6.04 (1.27) & 6.08 (0.91) &  \icc{.686} & 0.402 & 0.849 & 6.20 (1.15) & 6.24 (0.93) &  \icc{.686} & 0.402  & 0.849 & 5.84 (1.55) & 5.80 (1.41) &  \icc{.696} & 0.418  & 0.854 & 5.96 (1.57) & 6.12 (1.33) &  \icc{.023} & -0.387 & 0.415 \\
CG      & 5.92 (1.22) & 5.76 (1.33) &  \icc{.630} & 0.321 & 0.818 & 5.92 (1.41) & 6.12 (1.09) &  \icc{.429} & 0.048  & 0.701 & 5.40 (1.91) & 5.80 (1.68) &  \icc{.647} & 0.353  & 0.826 & 5.64 (1.85) & 5.40 (1.83) &  \icc{.702} & 0.435  & 0.856 \\
PF      & 5.32 (1.28) & 5.40 (1.41) &  \icc{.508} & 0.142 & 0.750 & 5.24 (1.45) & 4.84 (1.77) &  \icc{.670} & 0.389  & 0.839 & 4.80 (1.83) & 4.96 (1.84) &  \icc{.772} & 0.549  & 0.893 & 5.08 (1.87) & 4.80 (1.96) &  \icc{.610} & 0.294  & 0.807 \\
AUT     & 5.24 (1.54) & 5.48 (1.33) &  \icc{.471} & 0.103 & 0.727 & 5.32 (1.63) & 4.76 (1.51) &  \icc{.597} & 0.275  & 0.799 & 5.04 (1.43) & 4.80 (1.63) &  \icc{.711} & 0.451  & 0.861 & 5.68 (1.44) & 5.44 (1.33) &  \icc{.145} & -0.267 & 0.508 \\
CUR     & 5.96 (0.84) & 5.88 (1.17) &  \icc{.447} & 0.064 & 0.714 & 5.64 (1.44) & 5.68 (1.60) &  \icc{.426} & 0.034  & 0.701 & 5.28 (1.34) & 5.36 (1.41) &  \icc{.722} & 0.463  & 0.868 & 5.68 (1.22) & 5.84 (1.40) &  \icc{.555} & 0.211  & 0.776 \\
IMM     & 6.56 (0.58) & 6.24 (1.09) &  \icc{.399} & 0.035 & 0.676 & 5.96 (1.17) & 6.00 (1.22) &  \icc{.590} & 0.255  & 0.797 & 6.08 (1.38) & 5.80 (1.38) &  \icc{.097} & -0.310 & 0.471 & 6.16 (1.07) & 6.24 (0.97) &  \icc{.491} & 0.121  & 0.740 \\
MAS     & 5.76 (1.20) & 5.52 (1.23) &  \icc{.484} & 0.123 & 0.733 & 5.40 (1.35) & 5.64 (1.22) &  \icc{.493} & 0.133  & 0.739 & 5.08 (1.55) & 5.20 (1.22) &  \icc{.511} & 0.148  & 0.752 & 5.48 (1.58) & 5.36 (1.22) &  \icc{.313} & -0.096 & 0.629 \\
MEA     & 4.84 (1.31) & 5.00 (1.41) &  \icc{.698} & 0.426 & 0.854 & 4.76 (1.39) & 4.60 (1.78) &  \icc{.523} & 0.165  & 0.758 & 4.52 (1.90) & 4.52 (1.39) &  \icc{.692} & 0.411  & 0.852 & 4.76 (1.36) & 4.76 (1.56) &  \icc{.698} & 0.422  & 0.855 \\
ENJ     & 6.04 (0.84) & 6.20 (0.65) &  \icc{.647} & 0.353 & 0.826 & 5.84 (1.28) & 5.84 (1.55) &  \icc{.637} & 0.326  & 0.823 & 5.68 (1.57) & 5.64 (1.41) &  \icc{.811} & 0.615  & 0.912 & 5.96 (1.21) & 5.84 (1.25) &  \icc{.665} & 0.373  & 0.837 \\
\midrule
NPS     & 6.36 (2.22) & 6.64 (2.14) &  \icc{.827} & 0.650 & 0.920 & 5.72 (2.67) & 5.92 (2.81) &  \icc{.705} & 0.502  & 0.879 & 6.32 (3.08) & 6.68 (2.75) &  \icc{.871} & 0.733  & 0.941 & 6.44 (2.66) & 5.92 (3.23) &  \icc{.743} & 0.504  & 0.877 \\
\midrule
APR     & 5.80 (1.00) & 5.92 (0.64) &  \icc{.687} & 0.411 & 0.848 & 5.44 (1.53) & 5.60 (1.53) &  \icc{.725} & 0.469  & 0.868 & 5.52 (1.45) & 5.52 (1.48) &  \icc{.868} & 0.723  & 0.940 & 6.04 (1.02) & 5.88 (1.17) &  \icc{.867} & 0.724  & 0.939 \\
\bottomrule
\end{tabularx}
\begin{flushleft}
\footnotesize{$n = 25$ per group. AA - \textsc{Audiovisual Appeal}, CH - \textsc{Challenge}, EC - \textsc{Ease of Control}, CG - \textsc{Clarity of Goals}, PF - \textsc{Progress Feedback}, AUT - \textsc{Autonomy}, CUR - \textsc{Curiosity}, IMM - \textsc{Immersion}, MAS - \textsc{Mastery}, MEA - \textsc{Meaning}, ENJ - \textsc{Enjoyment}, NPS - \emph{Net Promoter Score}, APR - Appreciation
}
\end{flushleft}
\end{table}
\end{landscape}
}

\subsection{RQ1: How does the \emph{miniPXI} perform with respect to test-retest reliability?}

\subsubsection{Overall test-retest reliability}
The test-retest reliability of the \emph{miniPXI} constructs was evaluated through the computation of ICCs, with values ranging from .365 to .704, indicating poor to moderate reliability, see Table~\ref{table:all_questionnaires}. Notably, the constructs with the highest ICC values were \emph{Enjoyment} (ICC = .704, 95\% CI [.59, .791]), \textsc{Clarity of Goals} (ICC = .624, 95\% CI [.488, .731]), and \textsc{Progress Feedback} (ICC = .651, 95\% CI [.522, .751]), indicating moderate test-retest reliability. In contrast, constructs such as \emph{Immersion} (ICC = .365, 95\% CI [.183, .523]) and \textsc{Challenge} (ICC = .422, 95\% CI [.249, .57]) exhibited lower ICC values indicating poor reliability.

\subsubsection{Test-retest reliability across the games}
The test-retest reliability analysis of the \emph{miniPXI} showed varying findings across the different games as summarized in Table~\ref{table:minipxi_games}. 

In the first-person shooter game \emph{Station Saturn}, ICC values for the constructs of the \emph{miniPXI} ranged from .365 to .767, indicating poor to good reliability. The construct with the highest ICC value was \textsc{Autonomy}, i.e., .767 (95\% CI [.535, .892]), indicating good test-retest reliability. The construct of \textsc{Progress Feedback} exhibited moderate reliability, with an ICC value of .685 (95\% CI [.391, .851]). Conversely, the construct of \textsc{Meaning} demonstrated the lowest ICC value of .365 (95\% CI [-.05, .665]), followed by \textsc{Immersion}, which also exhibited relatively lower reliability, with an ICC value of .428 (95\% CI [.034, .706]). 

In the evaluation of \emph{Little Runmo}, the \emph{miniPXI} constructs showed varying levels of test-retest reliability. The construct of \textsc{Audiovisual Appeal} and \textsc{Enjoyment} stood out for their good reliability, with ICC values of .84 (95\% CI [0.671, 0.926]) and .809 (95\% CI [.614, .911], respectively. The remaining constructs had moderate reliability, with \textsc{Challenge} having an ICC of .534 (95\% CI [0.191, 0.762]), \textsc{Ease of Control} at .549 (95\% CI [0.214, 0.771]), and \textsc{Curiosity} at .638 (95\% CI [0.332, 0.823]. In contrast, \textsc{Autonomy} showed poor reliability with ICC values of .295 (95\% CI [-0.118, 0.617]), and \textsc{Immersion} demonstrated very poor reliability with ICC values of .238 (95\% CI [-0.173, 0.576]). 

In the study of \emph{Empty}, the values across the constructs were quite low compared to the other games. Only the construct of \textsc{Meaning} demonstrated good reliability, with an ICC value of .783 (95\% CI [0.565, 0.898]). \textsc{Enjoyment} and \textsc{Autonomy} demonstrated moderate reliability with ICC values of .656 (95\% CI [0.369, 0.828]) and .505 (95\% CI [0.155, 0.745]). \textsc{Challenge}, on the other hand, had very poor reliability, with an ICC of .014 (95\% CI [-0.33, 0.379]), indicating that participants' perceptions of the game's difficulty may vary over time. Furthermore, with an ICC of .043 (95\% CI [-0.324, 0.413]), \textsc{Mastery} also showed very poor reliability, implying fluctuations in participants' perceived sense of control and skill advancement in this puzzle game. 

In the case of \emph{Sort the Court}, the test-retest reliability of the \emph{miniPXI}'s constructs yielded diverse outcomes. The construct of \textsc{Enjoyment} demonstrated good reliability, with an ICC value of .796 (95\% CI [0.599, 0.903]). Furthermore, only three constructs, i.e., \textsc{Clarity of Goals}, \textsc{Progress Feedback}, and \textsc{Meaning}, showed moderate reliability with ICC values of .739 (95\% CI [0.499, 0.874]), .686 (95\% CI [0.418, 0.845]), and .598 (95\% CI [0.279, 0.798]). Conversely, \textsc{Ease of Control} and \textsc{Autonomy} displayed very poor test-retest reliability, with ICC values of .165 (95\% CI [-0.246, 0.518]) and 0.232 (95\% CI [-0.178, 0.568]).

\subsubsection{Test-retest reliability across the groups}
Similar to the individual game-based test-retest reliability analysis of the \emph{miniPXI}, mixed findings were observed across different groups (see Table~\ref{table:minipxi_subgroups}):

\begin{description}[labelindent=0.0\parindent,leftmargin=0.0\parindent,itemsep=1.0\baselineskip]
    \item[Group - PXI:] In the group where participants filled out the \emph{miniPXI} and the \emph{PXI} as an accompanying measure, seven out of 11 constructs demonstrated moderate reliability. Constructs such as \textsc{Meaning} and \textsc{Ease of Control}, showed moderate reliability, with ICC values of .698 (95\% CI [0.426 - 0.854]) and .686 (95\% CI [0.402 - 0.849]), respectively. However, constructs of \textsc{Immersion} and \textsc{Curiosity} showed poor reliability, with ICC values of .399 (95\% CI [0.035 - 0.676]) and .447 (95\% CI [0.064 - 0.714]), respectively. 
    
    \item[Group - PENS:] In the group where participants completed the \emph{PENS} measure in addition to the \emph{miniPXI}, seven out of 11 constructs demonstrated moderate reliability. The construct of \textsc{Audiovisual Appeal} had the highest ICC value of .73 (95\% CI [0.479, 0.871]), followed by \textsc{Ease of Control} with an ICC value of .686 (95\% CI [0.402, 0.849]). However, the construct of \textsc{Challenge} demonstrated very poor reliability, with an ICC value of .212 (95\% CI [-0.205, 0.559]). 
    
    \item[Group - GEQ:] In the group that used the \emph{miniPXI} and the \emph{GEQ} measure, nine out of 11 constructs demonstrated moderate to good test-retest reliability. The constructs of \textsc{Enjoyment}, \textsc{Progress Feedback}, and \textsc{Audiovisual Appeal} demonstrated good reliability, with ICC values of .811 (95\% CI [0.615, 0.912]), .772 (95\% CI [0.549, 0.893]), and .756 (95\% CI [0.523, 0.884]), respectively. However, the construct of \textsc{Immersion} demonstrated very poor reliability, with an ICC value of 0.097 (95\% CI [-0.31, 0.471]), indicating that participants' immersion levels may vary across sessions. 
    
    \item[Group - AttrakDiff:] In the group which included the \emph{miniPXI} and \emph{AttrakDiff}, the \emph{miniPXI} constructs demonstrated very poor to moderate test-retest reliability. Particularly, the constructs such as \textsc{Clarity of Goals} and \textsc{Meaning} were at the top, with ICC values of .702 (95\% CI [0.435, 0.856]) and .698 (95\% CI [0.422, 0.855]). In contrast, the constructs of \textsc{Ease of Control} and \textsc{Autonomy} demonstrated very poor test-retest reliability with low ICC values of .023 (95\% CI [-0.387, 0.415]) and .145 (95\% CI [-0.267, 0.508]).
\end{description}

\subsection{RQ2: How does the \emph{miniPXI} perform in test-retest reliability when compared to other popular measures that use multiple items per construct?}

To assess the test-retest reliabilities of the multi-item measures, the internal consistency of each of the included constructs, followed by the ICC values per construct per measure (see Table~\ref{table:all_questionnaires}), were calculated. 

\subsubsection{Internal Consistency} 
Cronbach's $\alpha$, and McDonald's $\omega$ for the \emph{PXI} were consistent and demonstrated high internal consistency for both week 1 ($\alpha = .71-.96, \omega = .79-.96$) and week 3 ($\alpha = .72-.94, \omega = .76-.94$). The constructs of \emph{PENS} also demonstrated high internal consistency for both week 1 ($\alpha = .77-.88, \omega = .74-.89$) and week 3 ($\alpha = .87-.93, \omega = .85-.94$), with the exception of the construct of \textsc{Intuitive Controls} where the value was slightly lower ($\alpha = .66, \omega = .74$) for week 1. For the \emph{GEQ}, except the week 1 value for the construct of \emph{Presence} ($\alpha = .66, \omega = .7$), all multi-item constructs showed high internal consistency for both week 1 ($\alpha = .79-.81, \omega = .70-.82$) and week 3 ($\alpha = .76-.82, \omega = .77-.83$). However, for the \emph{GEQ}, the reliability coefficients for the construct of \textsc{Immersion} were not calculated as only one item exists in that construct. \emph{AttrakDiff} also demonstrated high internal consistency for all constructs in week 1 ($\alpha = .71-.91, \omega = .74-.91$) and for three constructs other than \textsc{Pragmatic Quality} in week 3 ($\alpha = .82-.91, \omega = .83-.92$). However, the values for the construct of \textsc{Pragmatic Quality} were lower for week 3 ($\alpha = .57, \omega = .59$)

\begin{description}[labelindent=0.0\parindent,leftmargin=0.0\parindent,itemsep=0.5\baselineskip]
    \item[\emph{PXI}:] In assessing the test-retest reliability of the \emph{PXI}, all of the constructs, except \textsc{Curiosity}, demonstrated moderate to good test-retest reliability, with ICC values ranging from .566 to .881 and corresponding 95\% confidence intervals of [0.221, 0.784] to [0.749, 0.946]. The constructs of \textsc{Meaning}, \textsc{Clarity of Goals}, and \textsc{Challenge} had the highest values, showing good test-retest reliability, with ICC values exceeding .80, along with narrow 95\% confidence intervals of [0.634, 0.916] to [0.749, 0.946]. The construct of \textsc{Curiosity}, however, demonstrated poor reliability with an ICC value of 0.495 (95\% CI [0.137, 0.740]). 
    
    \item[\emph{PENS}:] For the \emph{PENS}, the ICC values across the constructs varied from .47 to .734, with the corresponding 95\% confidence intervals ranging from [0.093, 0.727] to [0.48, 0.874], demonstrating poor to moderate test-retest reliability. The constructs of \textsc{Presence/Immersion} and \textsc{Autonomy} had the highest ICC values of .734 (95\% CI [0.48, 0.874]) and .695 (95\% CI [0.422, 0.853]), showing moderate reliability. In contrast, the construct of \emph{Competence}, with an ICC value of .47 (95\% CI [0.093, 0.727]), is indicative of poor test-retest reliability. 
    
    \item[\emph{GEQ}:] The test-retest reliability analysis of the \emph{GEQ} showed good reliability for all constructs with ICC values ranging from .753 to .852, along with 95\% confidence intervals ranging from [0.514 to 0.883] to [.695 to .932]. Notably, constructs of \textsc{Absorption} and \textsc{Flow} had ICC values greater than .8 and confidence intervals of [0.684, 0.93] and [0.695, 0.932], respectively. Furthermore, the constructs of \textsc{Presence} and \textsc{Immersion} had ICC values of .765 (95\% CI [0.54, 0.889]) and .753 (95\% CI [0.514, 0.883]). 
    
    \item[\emph{AttrakDiff}:] The \emph{AttrakDiff} measure exhibited moderate to good test-retest reliability for all constructs, with ICC values ranging from .715 to .853, along with 95\% confidence intervals ranging from [0.457, 0.862] and [0.694, 0.932]. The constructs of \textsc{Hedonic Quality - Stimulation}, \textsc{Hedonic Quality - Identification}, and \textsc{Attractiveness}, demonstrated good reliability with ICC values of .75 (95\% CI [0.508, 0.882]), .811 (95\% CI [0.617, 0.912]), and .853 (95\% CI [0.694, 0.932]), respectively. Only the construct of \emph{Pragmatic Quality} demonstrated a moderate, yet still commendable, level of reliability, as indicated by an ICC value of .715 (95\% CI [0.457, 0.862]). 
\end{description}

\subsection{RQ3: How does the overall enjoyment item of the \emph{miniPXI} compare in test-retest reliability to the \emph{NPS}?}
To assess the test-retest reliabilities of the \emph{NPS} versus the enjoyment item, ICC values for each were calculated, as well as for the 'appreciation' item. 

\subsubsection{NPS}
Although a single-item measure, the adapted \emph{NPS} showed good test-retest reliability across the entire dataset, i.e., 100 responses, see Table~\ref{table:all_questionnaires}. The ICC value of .791 and a 95\% confidence interval of [0.704, 0.854] indicate that the respondents' inclination to recommend the game to others remained consistently steady over time, demonstrating the long-term viability of the \emph{NPS} as a measure for assessing user satisfaction and commitment in a gaming context. The average \emph{NPS} for the whole dataset was 6.21 ($SD = 2.65$) in week 1 and 6.29 ($SD = 2.74$) in week 3, indicating a small overall increase in the likelihood of a recommendation.

The test-reliability of the \emph{NPS}, when assessed for individual games, demonstrated moderate to good reliability (cf. Table~\ref{table:minipxi_games}). \emph{Little Runmo} demonstrated the highest test-retest reliability among the games, with an ICC value of .839 (95\% CI [0.664, 0.926]). Conversely, \emph{Sort the Court} indicated moderate reliability, with an ICC value of .729 (95\% CI [0.48, 0.869]). 

Moreover, analyzing \emph{NPS} scores by groups reveals moderate to good test-retest reliability, see Table~\ref{table:minipxi_subgroups}. The group \emph{GEQ} participants consistently provided the highest \emph{NPS} scores across both sessions, with a mean score of 6.32 ($SD = 3.08$) in week 1 and 6.68 ($SD = 2.75$) in week 3, with the highest ICC value of .871 (95\% CI [0.733, 0.941]). The \emph{NPS} demonstrated moderate reliability in only Group \emph{AttrakDiff} with an ICC value of .743 (95\% CI [0.502, 0.879]).

\subsubsection{Appreciation}
The 'appreciation' single-item, showed good test-retest reliability in the overall dataset (see Table~\ref{table:all_questionnaires}) with an ICC value of .795 and a 95\% CI of [0.71, 0.857]. The mean appreciation score was 5.70 ($SD = 1.28$) in week 1 and 5.73 ($SD = 1.25$) in week 3, indicating a slight improvement in overall appreciation between the two sessions.

An analysis of the appreciation item scores for each game reveals some variations in the consistency of the results across the sessions, see Table~\ref{table:minipxi_games}. The responses from \emph{Little Runmo} demonstrated a good level of test-retest reliability, the best among all the included games, as evidenced by an ICC value of .85 (95\% CI [0.689, 0.931]). This indicates that users consistently appreciated the game across both sessions. On the other hand, responses for \emph{Empty} exhibited a moderate level of test-retest reliability, as indicated by an ICC value of .66 (95\% CI [0.367, 0.837]).

Examining the appreciation scores across different groups demonstrated moderate to good reliability (cf. Table~\ref{table:minipxi_subgroups}). The scores in Group \emph{GEQ} and Group \emph{AttrakDiff} demonstrated good test-retest reliability with ICC values of .868 (95\% CI [0.723, 0.94]) and .867 (95\% CI [0.724, 0.939]). In contrast, participants in Group \emph{PXI} demonstrated a moderate but still commendable level of reliability, as indicated by an ICC value of .687 (95\% CI [0.411, 0.848]). 

\subsubsection{miniPXI overall Enjoyment}
The single overall enjoyment item of the \emph{miniPXI} demonstrated moderate to good test-retest reliability in the overall dataset, with an ICC value of .704 (95\% CI [0.59, 0.791]), indicating that users consistently experience similar levels of enjoyment during gaming sessions, see Table~\ref{table:all_questionnaires}. The average enjoyment score remained consistent over time, with a mean of 5.88 (SD = 1.24) in week 1 and 5.88 (SD = 1.26) in week 3.

Examining the overall enjoyment scores of individual games uncovers differences in the consistency of test-retest reliability, see Table~\ref{table:minipxi_games}. For \emph{Little Runmo}, the single enjoyment item showed good test-retest reliability, with an ICC value of .809 (95\% CI [0.614, 0.911]). For \emph{Station Saturn}, however, the single enjoyment exhibited moderate reliability, as indicated by a lower ICC value of .597 (95\% CI [0.26, 0.804]). 

Moreover, the overall enjoyment ratings based on different groups also showed moderate to good test-retest reliability, as shown in Table~\ref{table:minipxi_subgroups}. The responses from Group \emph{GEQ} demonstrated good reliability, with an ICC value of .811 (95\% CI [0.615, 0.912]). On the other hand, values from the other three groups demonstrated moderate reliability, with ICC values ranging from .637 (95\% CI [0.326, 0.823]) to .665 (95\% CI [0.373, 0.837]).

\section{Discussion}
    \label{sec:discussion}

This study assessed the test-retest reliability of the \emph{miniPXI} over a period of three weeks. The findings of this study suggest potential differences in test-retest reliability across different game genres. In the following, we first contrast the performance of the \emph{miniPXI} to the multi-item measures. Next, we reflect on the results obtained via single-item measurements to assess the overall 'appreciation' of the post-game experience of players. Lastly, we discuss the theoretical implications of our study with respect to the inherently dynamic nature of PX, and how this may impact test-retest reliability. We end this section with practical considerations for the use of the \emph{miniPXI} in the context of repeated measurements.

\subsection{Test-retest Reliability of the \emph{miniPXI} as Compared to Multi-item per Construct Measures}

In this study, we set out to explore whether test-retest reliability scores of the \emph{miniPXI} would still be considered within the parameters of what has been considered acceptable
~\cite{KOO2016155}. While there are no standard universal values for acceptable test-retest, generally, values $> .5$ for the ICC are suggested. Respecting this threshold, our study yields mixed results concerning the repeatability of the \emph{miniPXI} measurements. 

\subsubsection{Test-retest reliability of the \emph{miniPXI}}
The analysis of the \emph{miniPXI} test-retest reliability revealed varying results in performance. Based on the ICC values computed for the whole dataset of 100 responses, the \emph{miniPXI} showed moderate to poor test-retest reliability. Notably, the umbrella construct of \textsc{Enjoyment} demonstrates the highest reliability, with an ICC of .704. In contrast,  \textsc{Immersion}~\cite{Jennet_IEQ_2008} scored the lowest ICC of .365. The analysis of the test-retest reliability of the \emph{miniPXI} across different games resulted in similarly varied outcomes. However, considering relatively a small sample size per game  and genre necessitates caution in interpreting these results. The construct of \textsc{Enjoyment} scored consistently above .5. Interestingly, test-retest reliability scores were far better for the platformer game (nine out of 11 constructs scoring moderate or good) and the first-person shooter game (seven out of 11 constructs scoring moderate or good). 
For the puzzle game, nine out of 11 constructs showed poor to very poor reliability. Similarly, eight of the 11 constructs showed poor to very poor reliability for the simulation game. The analysis of the test-retest reliability of the \emph{miniPXI} across different groups (according to the multi-item measures) equally resulted in varied outcomes. Here, the ICC of \textsc{Enjoyment} once more scored consistently above .5. Other constructs' ICC values varied. The construct of \textsc{Immersion} stood out in a negative manner, showing ICC values below .5 for three out of four groups, and a similarly problematic performance was observed for the test-retest reliability of \textsc{Mastery}.

\subsubsection{Comparison with multi-item measures}
In contrast to the mixed results of the \emph{miniPXI}, the established multi-item measures used in this investigation, i.e., the \emph{PXI}, \emph{PENS}, \emph{GEQ}, and \emph{AttrakDiff}, overall demonstrated moderate to good test-retest reliability across the constructs.  The performance of the measures demonstrates their suitability for capturing the intricacies of psychological variables associated with PX through repeated measurements. 

Constructs within the \emph{PXI} demonstrated moderate to good reliability with only the construct of \textsc{Curiosity}, still scoring borderline (ICC = .495). As the \emph{PXI} is considered a parsimonious measure, with three items per dimension only, the test-retest reliability scores are reassuring on an overall level, and support its applicability to research designs that demand repeated evaluations. 

Likewise, the \emph{PENS} demonstrated moderate to good test-retest reliability in measuring \textsc{Autonomy} (3 items), \textsc{Relatedness} (3 items), \textsc{Presence/Immersion} (9 items), and \textsc{Intuitive Controls} (3 items)~\cite{Ryan_PENS_2006}. Here, the construct of \textsc{Competence} (3 items) did score slightly below what is deemed acceptable (ICC = .470). 

The \emph{GEQ} demonstrated good test-retest reliability across all its four constructs: \textsc{Absorption} (5 items), \textsc{Flow} (9 items), \textsc{Presence} (4 items), and \textsc{Immersion} (1 item), all scoring above .75. Of particular interest is the good test-retest reliability observed for the single item for \textsc{Immersion} (ICC = .753), which outperformed the conceptually related constructs from the other included measures, i.e., the \textsc{Presence/Immersion} construct in the \emph{PENS} (ICC = .734), measured via nine items. This is also in stark contrast with the low ICC value of the \emph{miniPXI} single item for \textsc{Immersion} (ICC = .097). Note that the same participants filled out both items, i.e., the \textsc{Immersion} items of the \emph{GEQ} and the \emph{miniPXI}, for the same game. When comparing the wording of the single item for the \emph{miniPXI}, i.e., \textit{"I was fully focused on the game"} and the single item for the \emph{GEQ}, i.e., \textit{"I really got into the game"}, this raises intriguing questions about the factors influencing test-retest reliability in single-item assessments. It may be that the formulation for the measurement of immersion within the \emph{GEQ} captures a more stable and reliable aspect of the gaming experience. Another explanatory factor may be the use of different Likert scale structures, i.e., a 7-point scale ranging from "Strongly disagree (-3)" to "Strongly agree (3)" (\emph{miniPXI}) versus a 5-point Likert scale ranging from "Not at all (1)" to "Yes (5)" (\emph{GEQ}). 
This high ICC score for the \textsc{Immersion} item of the \emph{GEQ} may prompt a reevaluation of our reservation towards the measurement of immersion via a single item. Nevertheless, questions may still be raised with respect to the effectiveness of these single-item---assessing whether players really 'got into the game---in capturing all nuances of \textsc{Immersion}. Immersion, in particular, is acknowledged as central to game enjoyment and PX~\cite{mekler_systematic_2014}, but equally as a multifaceted and intricate construct, often requiring multiple items for comprehensive measurement~\cite{Jennet_IEQ_2008,brown_grounded_2004,ermi2005fundamental}. In fact, the \emph{GEQ} itself~\cite{BROCKMYER_GEQ_2009} comprises four constructs, totaling 19 items, polling for different underlying dimensions, such as presence, flow, cognitive absorption, and immersion.  

Finally, the \emph{AttrakDiff} measure also demonstrated good test-retest reliability across its four constructs, (\textsc{Attraktiveness} (7 items), \textsc{Identification} (7 items), \textsc{Stimulation} (7 items), and \textsc{Pragmatic quality} (7 items), as indicated by the moderate to good ICC values. These test-retest reliability scores confirm the \emph{AttrakDiff}'s capacity to reliably measure the intended constructs in the context of longitudinal studies and repeated measurement.\\
\par

\noindent In summary, our study underscores the limitations of single-item measurements in test-retest reliability when it comes down to the measurement of complex, inherently multi-dimensional player experiences. While some constructs in some experimental settings (platformer games, first-person shooters) demonstrated acceptable test-retest reliability, other constructs showed notable shortcomings. In addition, both \textsc{Immersion} and \textsc{Mastery} exhibited lower ICC scores in the overall sample of 100 participants and consistently scored lower at the genre and individual group levels. These results reaffirm the lower reliability score compared to other constructs while designing and developing the \emph{miniPXI}~\cite{haider_minipxi_2022}. The lower reliability scores underscore the limitations of using single-item measures for constructs that are known to be multidimensional in nature, such as \textsc{Immersion}~\cite{csikszentmihalyi_flow:_1990,brown_grounded_2004,Jennet_IEQ_2008} or \textsc{Mastery}~\cite{csikszentmihalyi_flow:_1990,brown_grounded_2004}.

\subsection{Test-retest reliability of NPS and Appreciation Item}
An analysis of the test-retest reliability of the single-item umbrella items in the study provided interesting findings, specifically related to the \emph{NPS}, 'appreciation', and the \emph{miniPXI} overall enjoyment item. When evaluating across the whole dataset, these three single items showed a consistent trend of moderate to good test-retest reliability. The consistency of this trend remained evident even when examined for different game genres and with different measure groups, indicating the strength and consistency of these metrics in assessing PX.

Among these three items, the \emph{NPS} and the 'appreciation' item consistently scored in the `good' range, whereas the overall enjoyment item scored in the moderate range. These findings suggest that both the adapted \emph{NPS} and the 'appreciation' item can be used to capture players' overall satisfaction and tendency to recommend a game. However, we need to critically reflect upon the extent to which such items capture what we typically understand by PX: Mekler et al.~\cite{mekler_systematic_2014} conceptualize game enjoyment as the valence of the PX, which is still reflected in the \emph{miniPXI} enjoyment item \emph{"I had a good time playing this game"}. In contrast, \textit{appreciating} a game and/or the likelihood of \textit{recommending a game} may capture different aspects beyond the PX itself. For example, these items may also tap into aspects such as providing \textit{`good value for money'}, relating to specific interests or contexts that fall beyond the realm of PX (e.g., serving an educational goal), or expressing high-level appreciation for a game rather than the experience associated with a specific gaming session.

\subsection{The Inherent (In)stability of Player Experiences}

Player experience, as any user experience, is not only subjective, it is also inherently dynamic and context-dependent~\cite{effielaw2009,josefPX2016}: PX is fluid in nature, and is impacted by prior mood or external circumstances that need not be related to a game itself~\cite{cairns2014}. Thus, a key philosophical and methodological question arises: Whether or not a single player experience can be \textit{duplicated} in a way that is comparable to the original, particularly over longer periods of time. Therefore, analyzing the evolution of PX over time poses a distinct challenge rooted in the need to distinguish actual shifts in perception of a player experience from external influences. Hence, commonly suggested values for test-retest reliability---proposed in the context of measuring more stable constructs such as personality traits~\cite{KOO2016155} or workload~\cite{Youngblut1993}---need to be critically appraised in the context of PX. 

In our study, we also assessed the test-retest reliability of four established and validated measures, comprising 24 multi-item constructs. We found test-retest reliability scores for these constructs ranging from .470 to .881, with an average of .715 ($SD = .113$). Hence, on average, these values range within what is considered as acceptable test-retest reliabilities. In turn, this suggests that PX, while considered dynamic, can still be measured as a \textit{consistent} experience over a time span of three weeks. Nevertheless, we also remark that scores were overall `modest': 13 constructs achieved moderate scores, only nine constructs scored in the `good' range, with an ICC above .750, and none of the constructs scored in the excellent range ($> .900$), and two constructs showed borderline test-retest reliabilities, slightly below what is considered `acceptable'. Therefore, we also recognize that compared to other measures assessing more stable constructs, games user researchers may need to accept more modest test-retest reliabilities. In the end, the question remains to what extent PX can be considered a \textit{stable} construct. This also underscores the need for adequate effect sizes and sample sizes. 

Here, there is an opportunity for our research community to explore further whether and which aspects of player experience can be expected to remain stable over time (e.g., perceived ease of use after the onboarding period), and which ones may fluctuate (e.g., to which extent a person feels challenged by a game, or whether gameplay satisfies a player's need for curiosity).

\subsection{Practical Considerations for Using the \emph{miniPXI} with Repeated Measurement}

As noted by \citet{Hassenzahl2003}: \textit{``Designers face the challenge of translating ‘experiential problems’ into improvement requirements without knowing whether the issues identified in the test are simply due to individual user characteristics such as mood, for example''}. In this last part of the discussion, we revisit the different findings in the context of how they can be used by games user researchers in practice.

Our study found that the \emph{miniPXI} showed rather moderate reliability when used alone~\cite{haider_minipxi_2022}, hence, its usefulness in repeated measurement and longitudinal studies, such as test-retest analysis, warrants caution.

Therefore, we suggest games user researchers, first and foremost, to take a critical look at the game under scrutiny and the specific question they aim to answer. In particular, we suggest researchers to assess the game's genre as it may warrant multi-item measures. 

In case researchers want to compare PX across different game iterations, at a more global level, they may still consider using the \emph{miniPXI}. In case games user researchers are particularly interested in scrutinizing specific PX dimensions, they should critically reflect on the impact of repeated measurement on the specific PX dimension in relation to the game at hand. For example, if different prototypes of the game still provide identical challenges, the measurement may reflect the impact of waning novelty or training effects. Moreover, for particular complex constructs such as immersion and/or mastery, we advise against the use of the \emph{miniPXI} alone, and suggest employing dedicated multi-item measures instead. Games user researchers could, for example, revert to the full \emph{PXI}, or reinsert the additional items for particularly those constructs of interest. However, researchers may also decide to opt for the \emph{GEQ}~\cite{BROCKMYER_GEQ_2009}, which offers four dimensions related to immersion, all scoring good in test-retest reliability. If constructs such as curiosity or aesthetic appeal are of interest, or constructs at the functional level, researchers may choose to add the \emph{AttrakDiff}~\cite{hassenzahl_aesthetics_2008}. If researchers want to investigate mastery or intuitive controls, they may want to resort to the PENS~\cite{Ryan_PENS_2006}.  

Lastly, researchers may consider the use of the \emph{NPS} or the `appreciation' item, which both provide a reliable measurement. Yet, they should remain aware that this may not fully capture the actual PX. If researchers aim to measure overall game enjoyment, we also recommend using the game enjoyment item. Here, it is safe to use the \emph{miniPXI}.

\section{Limitations and Future Work}
    \label{sec:limitations}
    
There are different open questions that invite further exploration. One of the primary limitations of this study is the relatively small sample size (N=24/25/25/26) used to assess test-retest reliability across different game genres. While the overall ICC score based on 100 responses for the \emph{miniPXI} indicates the limitation of single-item measures used in gaming, genre-specific analyses are constrained by the limited number of participants in each genre category. However, the small sample size for the different genres may not fully capture the variability and nuances in test-retest reliability for each genre. Consequently, the findings should be interpreted accordingly, and further research with larger and more diverse samples is necessary to confirm these results and provide more robust conclusions.

The second limitation of our research is that we only looked at a single game for each genre. Future research should conduct a more detailed analysis of multiple games per genre to determine whether the intrinsic complexity of these genres necessitates the employment of multi-item measures to capture PX over time accurately. On the other hand, more investigation could shed light on whether the observed suboptimal stability is due to inherent constraints of the items of the \emph{miniPXI} themselves, which would call for either refinement or change.

Third, the reliance of our study on remote participants during gaming sessions limits the lack of control and confirmation of their level of engagement. Moreover, their surroundings when they were engaged in the session could also have had an impact on their responses~\cite{carrigy2010}, especially when only a single-item per construct is being used to evaluate their PX. Future research should thus consider performing in-person studies to control the influence of the surroundings and how long the participants engaged with the game.  

A fourth limitation is presented by the evaluation and comparison of multi-item measures. The use of different participant samples for different measures/games increases unpredictability and potential biases. However, the choice to \textit{not} provide all four multi-item measures, was made intentionally to minimize participant fatigue and ensure reasonable completion times. Yet, future research may utilize a consistent sample over numerous multi-item scales to resolve this problem~\cite{Guttman1945}. Moreover, to the best of our efforts, we could not find any studies explicitly exploring the test-retest reliability of multi-item instruments, i.e., GEQ, PENS, PXI, and Attrakdiff. In the future, it would be interesting to compare the test-retest reliabilities of these multi-item measures.

A fifth limitation is related to the observed high scores on the \emph{NPS}, which might be influenced by the fact that we used free browser-based games. This raises questions about whether the players' likelihood of recommending the game could be affected by its cost. A future study could explore this by comparing \emph{NPS} scores between free and paid versions of games, shedding light on the impact of game pricing on player recommendations of the game.

Lastly, the amount of time between evaluation sessions might have had an impact, as the ideal period for determining the dependability of test-retest results in the gaming environment is yet unknown (cf. Section~\ref{sec:relatedwork}). In HCI, \citet{zniak2021} proposed a period of 14 days, yet in other sectors (e.g., Health), such time frames may differ, even span several months. Although we chose to conduct our research over three weeks, the influence of this span, particularly with regard to single-item per construct measures, should be investigated further. Moreover, the possible impact of the learning effect on repeated evaluations, in conjunction with the time span, could further unpack how PX evolves over a certain period of time. In the future, research attempts should study this element in more detail to determine the best time frame for accurately collecting PX over an extended period.

\section{Conclusion}
    \label{sec:conclusion}
    
The importance of repeated measurement in evaluating PX cannot be understated in GUR, especially in iterative game evaluations. However, this requires lightweight measurement instruments that can be administered in short periods of time. To this end, we conducted an examination of the test-retest reliability of the \emph{miniPXI}. The outcomes of our study are varied. 

Several constructs indicated moderate reliability, while others provided poor results. This points towards the need for careful consideration while using single-item measures in repeated measurements. Additionally, our study confirmed the advantage of multi-item measures in terms of repeatability across sessions. This reinforces their efficacy in reliably capturing player experiences over extended periods. As an additional point of interest, the examination of \emph{NPS}, `appreciation', and overall enjoyment utilizing single items demonstrated moderate to good repeatability within the gaming setting. 

In sum, our results suggest that it is important for games user researchers to take into account the unique characteristics of the game and the player experiences that are under scrutiny when choosing or creating measuring instruments. While some constructs in some experimental settings (platformer games, first-person shooters) suggest acceptable test-retest reliability, other constructs showed notable shortcomings.
In addition, we found that test-retest reliability is subpar for constructs known to be multidimensional, such as immersion or mastery. Nevertheless, in spite of the limitations that were discovered, there is still a potential value of the \emph{miniPXI} as a single-item measure in identifying variations in iterative game development situations at the global level. 

\bibliographystyle{ACM-Reference-Format}
\bibliography{bibliography}

\appendix

\end{document}